\documentclass[%
reprint,
 amsmath,amssymb,
 aps,
]{revtex4-2}
\usepackage{graphicx}
\usepackage{dcolumn}
\usepackage{bm}
\usepackage[flushmargin,ragged]{footmisc} 

\usepackage{color}
\usepackage{orcidlink}
\usepackage{hyperref}
\hypersetup
{
hypertex=true,
colorlinks=true,
linkcolor=blue,
filecolor=blue,
urlcolor=blue,
citecolor=blue,
breaklinks=true
}

\begin{document}

\preprint{APS/123-QED}

\title{A High-Frequency Flux-Driven DC-SQUID Threshold Detector at the Single-Microwave-Photon Energy Scale: A Numerical Feasibility Study}
\author{M. Y. Wang\orcidlink{0009-0006-6105-950X}}
\affiliation{HergD collaboration and Information Quantum Technology Laboratory, School of Information Science and Technology, Southwest Jiaotong University, Chengdu 610031, China}
\author{Y. Q. Chai\orcidlink{0009-0007-6246-9102}}
\affiliation{HergD collaboration and Information Quantum Technology Laboratory, School of Information Science and Technology, Southwest Jiaotong University, Chengdu 610031, China}
\author{S. N. Wang\orcidlink{}}
\affiliation{HergD collaboration and Information Quantum Technology Laboratory, School of Information Science and Technology, Southwest Jiaotong University, Chengdu 610031, China}

\author{L. F. Wei\orcidlink{0000-0003-1533-1550}}
\email{lfwei@swjtu.edu.cn}
\affiliation{HergD collaboration and Information Quantum Technology Laboratory, School of Information Science and Technology, Southwest Jiaotong University, Chengdu 610031, China}
\date{\today}
\begin{abstract}
In our recent work (Phys. Rev. B \textbf{113}, 224503 (2026)), we showed numerically that rapidly sweeping the bias current of a current-biased Josephson junction threshold detector (CBJJ-JTD) drives it out of equilibrium. This substantially reduces the effect of thermal noise on microwave detection and may enable threshold detection near the energy of a single microwave photon. The scheme requires very low damping ($\beta \leq 10^{-4}$) and a very fast bias-current sweep ($v/\beta \geq 1$), both of which are difficult to achieve with available junctions and low-pass-filtered cryogenic measurement lines. Here we propose a more experimentally accessible nonequilibrium threshold detector based on a high-frequency flux-driven direct-current superconducting quantum interference device (DC-SQUID) and investigate it by numerically solving the resistively and capacitively shunted junction (RCSJ) equations. The applied flux drives a circulating supercurrent and sustains nonequilibrium phase dynamics throughout the bias-current sweep, removing the need for a fast sweep or for transient oscillations set by the initial phase. Using typical experimental device parameters, we calculate switching-current distributions (SCDs) with and without a microwave input for both driven and undriven SQUIDs. We quantify detection performance using receiver operating characteristic (ROC) curves and their area under the curve (AUC). The simulations show that high-frequency flux drive reduces the impact of thermal noise and enables statistical discrimination of weak microwave pulses at the single-photon energy scale. We also numerically estimate detector characteristics such as phase selectivity and dynamic range.
\end{abstract}
\maketitle
\section{INTRODUCTION}

Sensitive microwave detectors are important for wireless communications~\cite{deng2024joint}, nondestructive security screening~\cite{brinker2020review}, medical imaging~\cite{chandra2015opportunities}, electronic countermeasures~\cite{jia2024detection}, and searches for high-frequency gravitational waves~\cite{berlin2022detecting,zheng2018electromagnetic} and the electromagnetic signatures of ultralight dark matter~\cite{bradley2003microwave,lasenby2021parametrics,staggs2018recent,senanian2024microwave}. High-fidelity detection of weak signals at the energy scale of a single microwave photon remains a major challenge~\cite{il2016microwave,zou2016photonics,albertinale2021detecting}.
Thermal noise limits the sensitivity of conventional metal-antenna microwave receivers to about $10^{-21}~\rm W/\sqrt{Hz}$~\cite{chai2025measuring}. Several newer devices have been developed for weak microwave signals. Graphene self-switching diodes have a noise-equivalent power (NEP) of approximately $2.2\times10^{-9}~\mathrm{W}/\sqrt{\mathrm{Hz}}$~\cite{westlund2015graphene}. Rydberg-atom microwave superheterodyne receivers have achieved an electric-field sensitivity of approximately $55~\mathrm{nV\,cm^{-1}}/\sqrt{\mathrm{Hz}}$~\cite{jing2020atomic}.
Although these approaches improve weak-signal detection, reaching the single-microwave-photon energy scale remains challenging. Cryogenic operation reduces thermal noise and can substantially improve microwave detection sensitivity.

Among cryogenic detectors based on semiconductors, graphene, and superconducting devices~\cite{wong2017quantum,lee2020graphene,fong2012ultrasensitive,oripov2023superconducting,inomata2016single}, Josephson-junction (JJ) detectors offer strong nonlinearity and controllable phase dynamics~\cite{chen2011microwave,pankratov2025detection}. JJ-based microwave detection schemes fall broadly into two classes. In the first, a superconducting qubit registers a state change induced by a single microwave photon~\cite{kono2018quantum,balembois2024cyclically,braggio2025quantum}. Coupling the qubit to a microwave resonator typically converts the incident traveling photon into a standing-wave cavity mode, strengthening its interaction with the qubit. Under dispersive coupling, a cavity photon shifts the qubit transition frequency, which can be read out to detect the photon. The overall efficiency depends on photon injection, qubit decoherence, and state-readout fidelity. The frequency-matching requirements of the cavity and qubit typically constrain the operating bandwidth. The second class uses voltage switching in a current-biased JJ to detect weak microwave signals~\cite{govenius2016detection,chai2026broadband,he2025experimental}. A junction operated in this regime is termed a Josephson threshold detector (JTD). The bias current and the high-frequency current induced by the incident microwave signal can jointly switch the junction from the zero-voltage state to a finite-voltage state. In the tilted-washboard picture, this switch corresponds to the escape of the phase particle from a potential well. The observed voltage switch serves as the detection signal. This mechanism allows broadband operation because it does not require precise matching to a narrow cavity mode. The induced current also depends on signal strength or photon number, giving JTDs the potential for photon-number resolution (PNR).

Thermal noise limits JTD sensitivity even at millikelvin temperatures, making it difficult to reach the energy scale of a single microwave photon.
Further cooling offers limited practical scope for improving sensitivity at these operating temperatures.
We therefore seek to use the intrinsic nonlinear dynamics of the JTD to lessen the impact of thermal noise while preserving or enhancing the microwave response.
In earlier work~\cite{chai2026measuring}, we found that a rapid bias-current sweep places a JTD in a nonequilibrium phase regime that is less sensitive to thermal noise and more responsive to weak microwave signals, potentially reaching the single-photon energy scale. The method relies on transient oscillations set by the initial junction phase, which decay through internal dissipation. It therefore requires both a junction with very low damping (or a very high quality factor $Q$) and a very fast current sweep. Our feasibility analysis indicated that available JTDs did not meet the device parameters used in that study. In addition, copper-powder and resistor--capacitor (RC) low-pass filters installed on the bias line to suppress environmental noise can have cutoff frequencies as low as approximately $2~\mathrm{kHz}$~\cite{martinis1987experimental,nazhestkin2025enhancing}, which limits the achievable sweep rate.

These results motivate a way to sustain nonequilibrium operation with experimentally accessible damping and bias-current sweep rates.
We propose threshold detection of weak microwave signals using a high-frequency flux-pumped direct-current superconducting quantum interference device (DC-SQUID). The external flux drives a circulating current, continuously sustaining nonequilibrium phase dynamics throughout the bias-current sweep. This sustained motion replaces the dissipatively decaying oscillations initiated by the single junction's initial phase in our prior work~\cite{chai2026measuring}. The scheme can operate at moderate sweep rates and damping.
Separate lines carry the high-frequency flux drive and the current sweep, allowing their filters and passbands to be designed independently. This arrangement makes it easier to exploit nonequilibrium phase dynamics experimentally.

Section II develops the resistively and capacitively shunted junction (RCSJ) model~\cite{10.1063/1.1651991} for the flux-driven DC-SQUID. Using common-mode and differential-mode phases, we examine how the drive modulates the instantaneous potential and phase dynamics, characterize nonequilibrium motion, and describe how to calculate switching-current distributions (SCDs) and quantify their separability with and without a microwave signal. Section III examines the effects of DC flux bias, the amplitude and frequency of the high-frequency flux drive, and thermal noise, then evaluates sensitivity to continuous-wave and pulsed microwave signals under optimized drive conditions. Section IV summarizes the results and discusses experimental feasibility and open questions.

\section{Model and Methods}
We model the phase dynamics of a flux-driven DC-SQUID and describe how we use simulated switching-current distributions to quantify microwave detection sensitivity.

\subsection{Phase dynamics of a DC-SQUID under external flux drive}
\begin{figure}[htbp]
\centering
\includegraphics[width=0.8\linewidth]{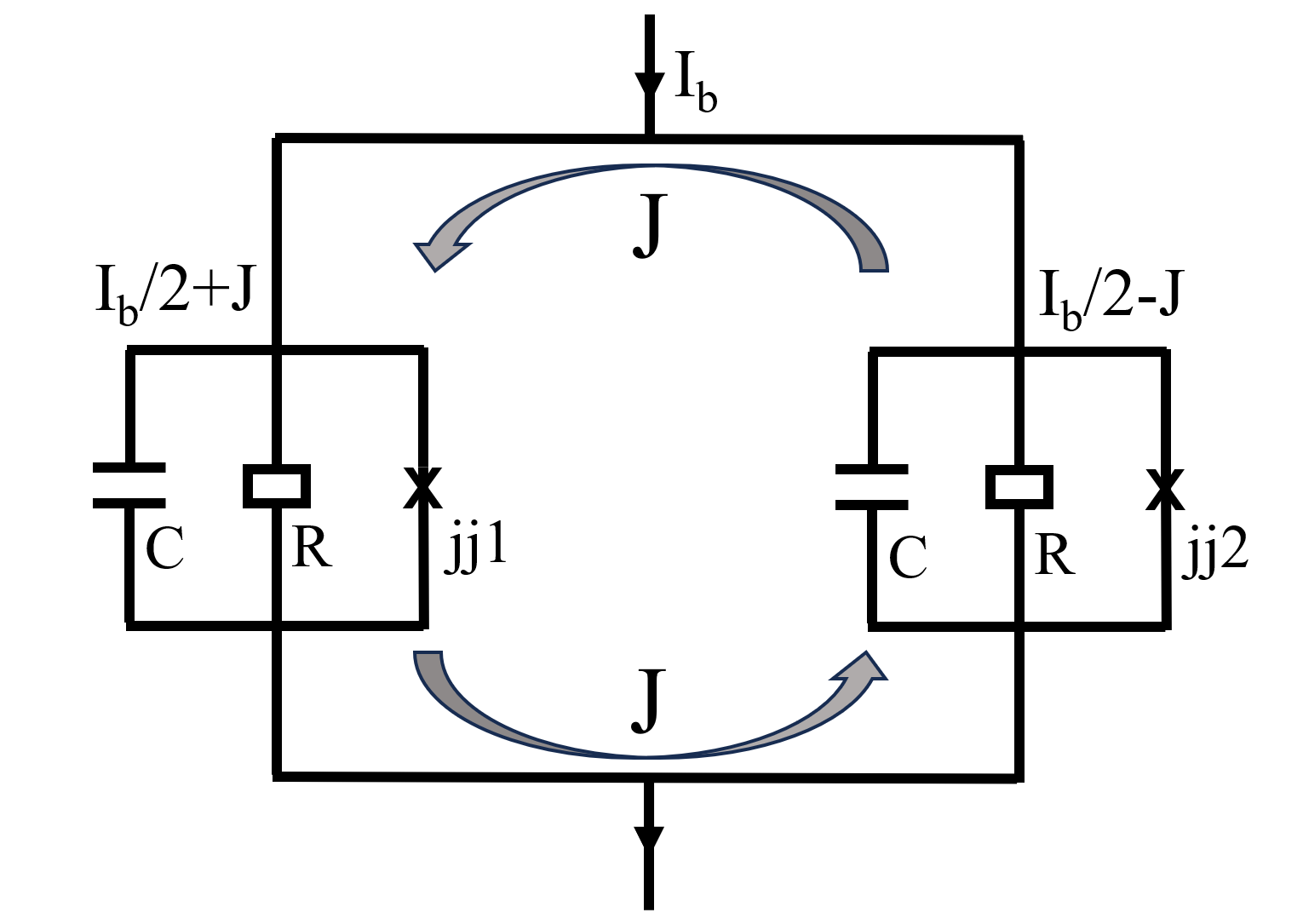}
\caption{RCSJ equivalent circuit of a symmetric DC-SQUID. $J$ denotes the circulating supercurrent in the loop, and $I_b$ is the applied bias current.}
\label{fig:RCSJ-SQUID}
\end{figure}

We consider a symmetric DC-SQUID comprising two identical Josephson junctions and neglect additional mutual inductive coupling outside the superconducting loop. Figure~\ref{fig:RCSJ-SQUID} shows the equivalent circuit. Each junction consists of an ideal Josephson element of critical current $I_0$ shunted by a resistor $R$ and a capacitor $C$. The total loop self-inductance is $L$, and the phase differences across the two junctions are $\varphi_1$ and $\varphi_2$.
With the sign convention for $J$ shown in Fig.~\ref{fig:RCSJ-SQUID}, the currents in the two branches under an applied bias $I_b$ are
\begin{equation}
\begin{cases}
I_1=\displaystyle\frac{I_b}{2}+J+I_{n1},\\[6pt]
I_2=\displaystyle\frac{I_b}{2}-J+I_{n2}.
\end{cases}
\label{eq:branch-current}
\end{equation}
Here $I_{n1}$ and $I_{n2}$ are independent effective thermal-noise currents. The flux quantization condition for the superconducting loop is~\cite{doi:https://doi.org/10.1002/3527603646.ch2}
\begin{equation}
\varphi_1-\varphi_2+
\frac{2\pi L J}{\Phi_0}
=
\frac{2\pi\Phi_{\mathrm{ext}}(t)}{\Phi_0},
\label{eq:flux-quantization}
\end{equation}
which gives the circulating current
\begin{equation}
J=
\frac{\Phi_0}{2\pi L}
\left[
\varphi_2-\varphi_1+
\frac{2\pi\Phi_{\mathrm{ext}}(t)}{\Phi_0}
\right].
\label{eq:circulating-current}
\end{equation}
Here $\Phi_0$ is the flux quantum and $\Phi_{\mathrm{ext}}(t)$ is the applied flux.
The RCSJ equations for the two junction phases are
\begin{equation}
\begin{cases}
\displaystyle
\frac{\Phi_0C}{2\pi}\frac{d^2\varphi_1}{dt^2}+\frac{\Phi_0}{2\pi R}\frac{d\varphi_1}{dt}+I_0\sin\varphi_1=
\frac{I_b}{2}+J+I_{n1},
\\[12pt]
\displaystyle
\frac{\Phi_0C}{2\pi}\frac{d^2\varphi_2}{dt^2}+
\frac{\Phi_0}{2\pi R}\frac{d\varphi_2}{dt}+
I_0\sin\varphi_2=\frac{I_b}{2}-J+I_{n2}.
\end{cases}
\label{eq:RCSJ-dimensional}
\end{equation}
The circulating current $J$ couples the junctions and is driven by the external flux $\Phi_{\mathrm{ext}}(t)$, which we write as
\begin{equation}
\Phi_{\mathrm{ext}}(t)
=\Phi_{\mathrm{dc}}+\Phi_{\mathrm{ac}}
\sin\left(\omega_{\mathrm{ac}}t+\psi_{\mathrm{ac}}
\right),
\label{eq:external-flux}
\end{equation}
Here $\Phi_{\mathrm{dc}}$ is the static flux bias, and $\Phi_{\mathrm{ac}}$, $\omega_{\mathrm{ac}}$, and $\psi_{\mathrm{ac}}$ are the amplitude, angular frequency, and initial phase of the high-frequency flux pump. The incident microwave signal induces an additional current, giving the total bias current
\begin{equation}
I_b(t)=v_I t+I_{\mathrm{rf}}
\sin\left(\omega_{\mathrm{rf}}t+\psi_{\mathrm{rf}}
\right),
\label{eq:bias-current}
\end{equation}
Here $v_I=dI_{\mathrm{dc}}/dt$ is the DC sweep rate, and $I_{\mathrm{rf}}$, $\omega_{\mathrm{rf}}$, and $\psi_{\mathrm{rf}}$ are the amplitude, angular frequency, and initial phase of the microwave current. Equations~\eqref{eq:RCSJ-dimensional}--\eqref{eq:bias-current} describe how the external flux, the microwave signal, and the bias current affect the SQUID dynamics.

We solve the coupled equations numerically. For this purpose, we define dimensionless time, current, and flux as
\begin{equation*}
\tau=\omega_Jt,\qquad
i_b=\frac{I_b}{I_0},\qquad
f=\frac{\Phi_{\mathrm{ext}}}{\Phi_0},
\label{eq:normalization-1}
\end{equation*}
and rewrite Eq.~\eqref{eq:RCSJ-dimensional} as
\begin{equation}
\begin{cases}
\begin{aligned}
\displaystyle
\frac{d^2\varphi_1}{d\tau^2}+
\beta\frac{d\varphi_1}{d\tau}+
\sin\varphi_1={}&
\frac{i_b(\tau)}{2}+\frac{\varphi_2-\varphi_1+2\pi f(\tau)}
{\pi\beta_L}
\\
&+
i_{n1}(\tau),
\end{aligned}
\\[12pt]
\begin{aligned}
\displaystyle
\frac{d^2\varphi_2}{d\tau^2}+\beta\frac{d\varphi_2}{d\tau}+\sin\varphi_2={}&
\frac{i_b(\tau)}{2}-\frac{\varphi_2-\varphi_1+2\pi f(\tau)}{\pi\beta_L}
\\
&+i_{n2}(\tau).
\end{aligned}
\end{cases}
\label{eq:RCSJ-normalized}
\end{equation}
Here $\omega_J=\sqrt{2eI_0/(\hbar C)}$ is the single-junction plasma frequency, $\beta=1/(RC\omega_J)$ is the dimensionless damping parameter, and $\beta_L=2LI_0/\Phi_0$ is the loop screening parameter. The normalized thermal-noise currents are $i_{nj}=I_{nj}/I_0$ for $j=1,2$.
The normalized applied flux and total bias current are
\begin{equation}
f(\tau)=f_{\mathrm{dc}}+a_{\mathrm{ac}}
\sin\left(\Omega_{\mathrm{ac}}\tau+\psi_{\mathrm{ac}}\right),
\label{eq:normalized-flux}
\end{equation}
\begin{equation}
i_b(\tau)=v\tau+A\sin\left(
\Omega_{\mathrm{rf}}\tau+\psi_{\mathrm{rf}}
\right),
\label{eq:normalized-current}
\end{equation}
The normalized parameters are $f_{\mathrm{dc}}=\Phi_{\mathrm{dc}}/\Phi_0$, $a_{\mathrm{ac}}=\Phi_{\mathrm{ac}}/\Phi_0$, $\Omega_{\mathrm{ac}}=\omega_{\mathrm{ac}}/\omega_J$, $A=I_{\mathrm{rf}}/I_0$, $\Omega_{\mathrm{rf}}=\omega_{\mathrm{rf}}/\omega_J$, and $v=v_I/(I_0\omega_J)$. We model the junction thermal-noise currents as Gaussian white noise, such that their normalized forms satisfy
\begin{equation}
\begin{aligned}
\left\langle i_{nj}(\tau)\right\rangle &=0,\\
\left\langle
i_{nj}(\tau)i_{nk}(\tau')
\right\rangle
&=2\beta D\,\delta_{jk}\delta(\tau-\tau'),
\label{eq:noise-mean}
\end{aligned}
\end{equation}
Here $D=k_BT/E_{J0}$ is the normalized thermal-noise intensity and
$E_{J0}=\hbar I_0/(2e)$ is the single-junction Josephson coupling energy.
\subsection{Nonequilibrium phase dynamics under sustained flux drive}
\label{subsec:nonequilibrium-definition}

The voltage state of a JJ can be described in terms of the motion of a phase particle in a tilted potential. In a DC-SQUID, the two phase degrees of freedom $\varphi_1$ and $\varphi_2$ give a two-dimensional potential. Writing the two junction equations as
\begin{equation}
\frac{\Phi_0 C}{2\pi}\ddot\varphi_j+
\frac{\Phi_0}{2\pi R}\dot\varphi_j+
\frac{2\pi}{\Phi_0}\frac{\partial U}{\partial\varphi_j}
=I_{nj},\qquad j=1,2,
\label{eq:potential-gradient-dynamics}
\end{equation}
and substituting Eq.~\eqref{eq:circulating-current}, we obtain the DC-SQUID potential
\begin{equation}
\begin{split}
U(\varphi_1,\varphi_2,t)={}&-E_{J0}\left(\cos\varphi_1+\cos\varphi_2\right)
\\
&-\frac{\Phi_0 I_b(t)}{4\pi}\left(\varphi_1+\varphi_2\right)\\
&+\frac{\Phi_0^2}{8\pi^2L}\left[\varphi_1-\varphi_2
-\frac{2\pi\Phi_{\mathrm{ext}}(t)}{\Phi_0}\right]^2,
\end{split}
\label{eq:potential-2D}
\end{equation}
Here \(E_{J0}=\Phi_0 I_0/(2\pi)\) is the Josephson coupling energy of a single junction. The three terms represent the Josephson coupling energy, the tilt due to the bias current, and the magnetic energy of the loop. The bias current drives both phases, while flux quantization relates their difference to the circulating current. To simplify the analysis, we introduce the common-mode phase \(\phi\) and the differential-mode phase \(\theta\)~\cite{doi:https://doi.org/10.1002/3527603646.ch2,kamal2012gain},
\begin{equation*}
\phi=\frac{\varphi_1+\varphi_2}{2},\qquad
\theta=\frac{\varphi_1-\varphi_2}{2},
\label{eq:phi-theta-definition}
\end{equation*}
where \(\phi\) describes the in-phase motion of the two junctions and
\(\theta\), half their phase difference, describes the antisymmetric motion.
The differential-mode phase and circulating current \(J\) are related by
\begin{equation}
\theta-\pi f(t)=-\frac{\pi L}{\Phi_0}J,
\label{eq:half-phase-difference-circulating-current}
\end{equation}
where \(f=\Phi_{\mathrm{ext}}/\Phi_0\) is the normalized applied flux and \(L\) is the loop self-inductance.
Using the trigonometric identity
\(\cos\varphi_1+\cos\varphi_2=2\cos\phi\cos\theta\),
together with \(\varphi_1+\varphi_2=2\phi\) and
\(\varphi_1-\varphi_2-2\pi f=2(\theta-\pi f)\), we rewrite the two-dimensional potential in Eq.~\eqref{eq:potential-2D} as
\begin{equation}
\begin{split}
U(\phi,\theta,t)={}&-2E_{J0}\cos\phi
\cos\theta-\frac{\Phi_0 I_b(t)}{2\pi}\phi
\\
&+\frac{\Phi_0^2}{2\pi^2L}\left[\theta-\pi f(t)\right]^2.
\end{split}
\label{eq:potential-phi-theta}
\end{equation}
Through the magnetic energy term involving \(\theta-\pi f(t)\), the applied flux shifts the preferred differential-mode phase over time. The resulting variation in $\theta$ changes the depth of the wells along the common-mode direction via the Josephson term \(-2E_{J0}\cos\phi\cos\theta\), periodically modulating the phase-particle escape rate.

In a single-junction JTD, nonequilibrium phase dynamics occurs when oscillations within the tilted potential well persist until the particle escapes~\cite{chai2026measuring}. These oscillations affect the switching-current distribution and reduce the sensitivity to thermal noise, improving microwave detection sensitivity.
The oscillations decay because of internal damping $\beta$. A single JJ therefore requires sufficiently low damping and a sufficiently large sweep rate $v$ for oscillations to persist until escape. The condition $v>\beta$ serves as a criterion for nonequilibrium operation in that scheme. These demanding device and sweep parameters are difficult to realize experimentally. The DC-SQUID offers a separate flux control in addition to current bias. A sustained flux drive replenishes energy lost to dissipation and can maintain phase oscillations throughout the sweep.

To choose suitable drive parameters and establish a quantitative criterion for nonequilibrium operation, we examine the deterministic trajectory under current bias and applied flux. Using the identities
\(\sin(\phi+\theta)+\sin(\phi-\theta)=2\sin\phi\cos\theta\)
and
\(\sin(\phi+\theta)-\sin(\phi-\theta)=2\cos\phi\sin\theta\), we obtain the common-mode and differential-mode phase equations
\begin{equation}
\begin{aligned}
&\frac{d^2\phi}{d\tau^2}+\beta\frac{d\phi}{d\tau}+\sin\phi\cos\theta=\frac{i_b(\tau)}{2}+\eta_\phi(\tau),\\
&\frac{d^2\theta}{d\tau^2}+\beta\frac{d\theta}{d\tau}
+\cos\phi\sin\theta\\
&\qquad=-\frac{2}{\pi\beta_L}\bigl[\theta-\pi f(\tau)\bigr]
+\eta_\theta(\tau),
\label{eq:mean-phase-equation}
\end{aligned}
\end{equation}
The corresponding thermal-noise terms are $\eta_\phi=(i_{n1}+i_{n2})/2$ and $\eta_\theta=(i_{n1}-i_{n2})/2$. Equation~\eqref{eq:noise-mean} then gives
\begin{equation}
\begin{aligned}
\left\langle \eta_\phi(\tau)\eta_\phi(\tau')\right\rangle
&=\beta D\,\delta(\tau-\tau'),\\
\left\langle \eta_\theta(\tau)\eta_\theta(\tau')\right\rangle
&=\beta D\,\delta(\tau-\tau'),\\
\left\langle \eta_\phi(\tau)\eta_\theta(\tau')\right\rangle
&=0.
\end{aligned}
\label{eq:transformed-noise-correlation}
\end{equation}
Equation~\eqref{eq:mean-phase-equation} shows that the applied flux acts directly on the differential-mode phase $\theta$ with a coefficient $2/\beta_L$. It modulates the common-mode phase $\phi$ indirectly through the
\(\sin\phi\cos\theta\) term. This coupling allows the flux drive to control the phase escape dynamics.

We next use Eq.~\eqref{eq:mean-phase-equation} to quantify departures from equilibrium. We omit thermal noise in this analysis to isolate the deterministic response.
In the nonequilibrium regime, the phase particle oscillates within the potential well and departs from the instantaneous well minimum. With the applied current $i_b(\tau)$ and flux $f(\tau)$, we quantify this departure by the distance between the actual phase coordinates $[\phi_{\rm det}(\tau),\theta_{\rm det}(\tau)]$ and the instantaneous minimum $[\phi_{m}(\tau),\theta_{m}(\tau)]$,
\begin{equation}
\mathcal D_{\mathrm{neq}}(\tau)
=\left|\phi_{\mathrm{det}}(\tau)-\phi_m(\tau)\right|
+\left|\theta_{\mathrm{det}}(\tau)-\theta_m(\tau)\right|
\label{eq:D-neq-instantaneous}
\end{equation}
which serves as a measure of nonequilibrium phase dynamics. $\mathcal D_{\mathrm{neq}}(\tau)$ includes displacements in both modes. We obtain the deterministic trajectory $[\phi_{\rm det}(\tau),\theta_{\rm det}(\tau)]$ by numerically solving Eq.~\eqref{eq:mean-phase-equation}.
Locating the instantaneous well minimum requires evaluating the two-dimensional potential in Eq.~\eqref{eq:potential-phi-theta} at each time $\tau$ for the current $i_b(\tau)$ and flux $f(\tau)$. The first derivatives,
\begin{equation}
\begin{aligned}
\frac{\partial U}{\partial\phi}
&=2E_{J0}\sin\phi\cos\theta
-\frac{\Phi_0 I_b}{2\pi},\\
\frac{\partial U}{\partial\theta}
&=2E_{J0}\cos\phi\sin\theta
+\frac{\Phi_0^2}{\pi^2L}
\left[\theta-\pi f(\tau)\right].
\label{eq:potential-phi-derivative}
\end{aligned}
\end{equation}
vanish at a local minimum. Setting both derivatives to zero and dividing by \(2E_{J0}=\Phi_0I_0/\pi\) gives the normalized stationary-point equations for $[\phi_{m}(\tau),\theta_{m}(\tau)]$,
\begin{equation}
\begin{aligned}
&\sin\phi_m\cos\theta_m=i_b(\tau)/2,\\
&\cos\phi_m\sin\theta_m
+\frac{2}{\pi\beta_L}[\theta_m-\pi f(\tau)]=0.
\label{eq:instantaneous-minimum-phi}
\end{aligned}
\end{equation}
The Hessian of the two-dimensional potential is
\begin{equation}
H_m=
\begin{pmatrix}
2E_{J0}\cos\phi_m\cos\theta_m &
-2E_{J0}\sin\phi_m\sin\theta_m\\[4pt]
-2E_{J0}\sin\phi_m\sin\theta_m &
2E_{J0}\cos\phi_m\cos\theta_m
+\dfrac{\Phi_0^2}{\pi^2L}
\end{pmatrix}
\label{eq:instantaneous-hessian}
\end{equation}
A solution of the stationary-point equations gives an instantaneous local minimum when both Hessian eigenvalues are positive.

\begin{figure}[htbp]
\centering
\includegraphics[width=0.95\linewidth]{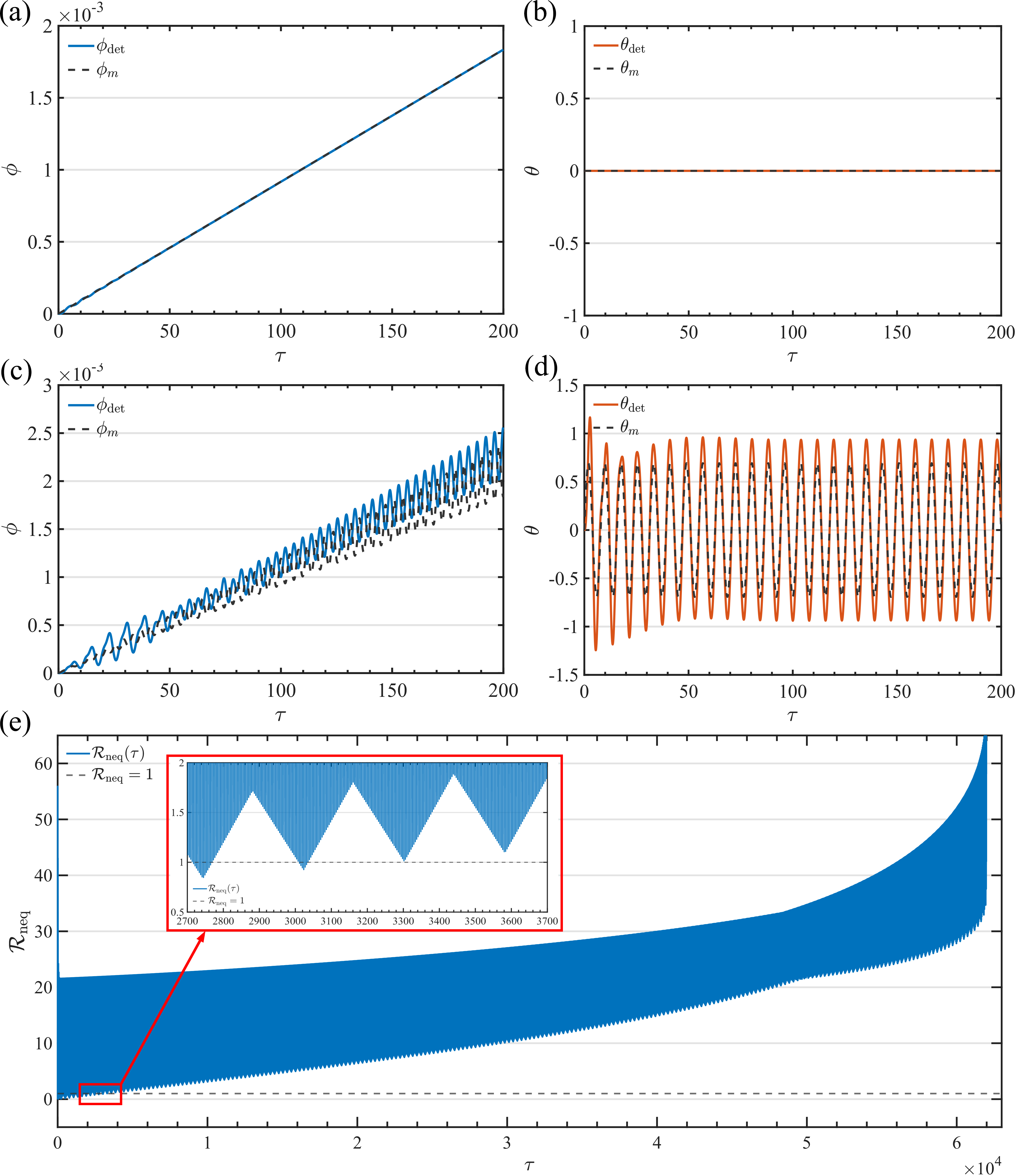}
\caption{Time evolution of the phase-particle coordinates $[\phi_{\rm det}(\tau),\theta_{\rm det}(\tau)]$ and the instantaneous well minimum $[\phi_{m}(\tau),\theta_{m}(\tau)]$ (a,b) without flux drive and (c,d) with flux drive. (e) Normalized displacement $\mathcal R_{\mathrm{neq}}(\tau)$ calculated from these trajectories using Eqs.~\eqref{eq:D-neq-instantaneous} and \eqref{eq:normalized-nonequilibrium-deviation}. The horizontal axes show dimensionless time $\tau=\omega_J t$, so the interval from the start of the bias sweep to the end of the trace in (e), at $\tau\simeq6.2\times10^4$, corresponds to approximately $3.0~\mu\mathrm{s}$. The bias-current sweep rate is $v=1.8355\times10^{-5}$. The flux parameters are $a_{\mathrm{ac}}=0.35$, $f_{dc}=0$, $\Omega_{ac}=0.8012$, and $\psi_{ac}=0$. The dimensionless integration step used throughout is $h=0.02$~\cite{arad2016g}.}
\label{fig:D-neq}
\end{figure}

We first calculate the displacement from the local minimum using Eq.~\eqref{eq:D-neq-instantaneous} without flux drive. During a slow bias-current sweep, the particle generally follows the moving well minimum, so the displacement remains small, as shown in Figs.~\ref{fig:D-neq}(a) and \ref{fig:D-neq}(b).
We use the maximum displacement under current bias alone as the reference scale for identifying nonequilibrium dynamics,
\begin{equation}
\mathcal D_0^{\mathrm{pk}}
=
\max_{0\leq\tau<\tau_{\mathrm{sw}}^{(0)}}
\mathcal D_{\mathrm{neq}}^{(0)}(\tau),
\label{eq:reference-deviation}
\end{equation}
where the superscript \(0\) denotes the absence of flux drive and \(\tau_{\mathrm{sw}}^{(0)}\) is the escape time.
Under flux drive, a displacement $\mathcal D_{\mathrm{neq}}^{(1)}(\tau)$ exceeding $\mathcal D_0^{\mathrm{pk}}$ signals departure from the undriven response. We define the normalized displacement $\mathcal R_{\mathrm{neq}}(\tau)$ to distinguish the phase-dynamical regimes,
\begin{equation}
\mathcal R_{\mathrm{neq}}(\tau)
=\frac{\mathcal D_{\mathrm{neq}}^{1}(\tau)}{\mathcal D_0^{\mathrm{pk}}}
\begin{cases}
>1,\text{Non-equilibrium state} \\
\leq1,\text{Equilibrium state}
\end{cases}.
\label{eq:normalized-nonequilibrium-deviation}
\end{equation}
Values $\mathcal R_{\mathrm{neq}}(\tau)>1$ indicate nonequilibrium dynamics, while $\mathcal R_{\mathrm{neq}}(\tau)\leq1$ corresponds to the near-equilibrium regime.

Figure~\ref{fig:D-neq} illustrates the criterion using numerical solutions of Eqs.~\eqref{eq:mean-phase-equation} and \eqref{eq:potential-phi-derivative}--\eqref{eq:instantaneous-hessian}. Figures~\ref{fig:D-neq}(a,b) and \ref{fig:D-neq}(c,d) show the trajectories of the phase particle and well minimum without and with flux drive, respectively. At the start of the sweep the particle is at the minimum, so \(\mathcal D_{\mathrm{neq}}(0)=0\). Without flux drive, the well evolves slowly and the particle tracks its minimum, keeping the displacement within the small scale set by \(\mathcal D_0^{\mathrm{pk}}\). With sustained high-frequency flux drive, the potential oscillates and the particle lags the moving minimum. The displacement \(\mathcal D^1_{\mathrm{neq}}(\tau)\) oscillates persistently with increasing amplitude, reaching the nonequilibrium regime defined by Eq.~\eqref{eq:normalized-nonequilibrium-deviation}, as shown in Fig.~\ref{fig:D-neq}(e).
\subsection{Numerical simulations and statistical analysis}
\label{subsec:numerical-method}
Having characterized the conditions for nonequilibrium phase dynamics, we now describe how we solve Eq.~\eqref{eq:RCSJ-normalized} and analyze its solutions to quantify microwave detection sensitivity.

As in our single-junction studies~\cite{chai2025measuring,chai2026measuring}, we detect weak microwave signals by distinguishing the SCDs with and without a microwave input. We repeatedly integrate Eq.~\eqref{eq:RCSJ-normalized} under each condition using a stochastic Heun predictor--corrector algorithm~\cite{heun1900neue,honeycutt1992stochastic}.
The state variables are the two junction phases and their time derivatives, $(\varphi_1,\dot\varphi_1,\varphi_2,\dot\varphi_2)$,
yielding four first-order stochastic differential equations. Within each integration step, we use the same noise sample for a given junction in the predictor and corrector stages. The junction noise currents are
\begin{equation}
i_{nj}^{(m)}=\sqrt{\frac{2\beta D}{h}}\,
\mathcal N_j^{(m)}(0,1),\qquad j=1,2,
\label{eq:discrete-noise}
\end{equation}
Here $\mathcal N_j^{(m)}(0,1)$ is a standard Gaussian variate with mean zero and unit variance for junction $j$ at time step $m$. The samples are independent across junctions and steps. We use experimentally representative parameters $I_0=140~\mathrm{nA}$, $R=480~\Omega$, $C=1~\mathrm{pF}$, and $L=3~\mathrm{nH}$~\cite{he2025experimental}, corresponding to $\beta\simeq0.101$ and $\beta_L\simeq0.406$. The bath temperature is $T=30~\mathrm{mK}$.

The DC bias starts at zero and increases at $v_I=0.053~\mathrm{A/s}$. We apply the flux drive and continuous-wave microwave current according to Eqs.~\eqref{eq:normalized-flux} and \eqref{eq:normalized-current}, respectively. The microwave input power is calibrated at a reference port with impedance $Z_0=50~\Omega$,
\begin{equation}
P_{\mathrm{rf}}=\frac{Z_0I_{\mathrm{rf}}^2}{2},\qquad
A=\frac{1}{I_0}\sqrt{\frac{2P_{\mathrm{rf}}}{Z_0}}.
\label{eq:input-power-calibration}
\end{equation}
Here $I_{\mathrm{rf}}$ is the peak total input current, $P_{\mathrm{rf}}$ is expressed in watts, and $P_{\mathrm{dBm}}=10\log_{10}[P_{\mathrm{rf}}/(1~\mathrm{mW})]$. We assume that the full current enters the SQUID and neglect line loss and impedance mismatch. The calibrated port power is distinct from the power dissipated in the junction shunt resistors. Section~\ref{subsec:pulsed-photon-response} describes the pulse energy calibration and injection timing. We independently sample the initial junction phases from $[-0.1,0.1]$ and set both initial phase velocities to zero, corresponding to an initially superconducting device. When flux is applied, we add independent perturbations sampled from $[-0.01,0.01]$ to the zero-current equilibrium phases corresponding to the DC component of the applied flux.

Following the numerical switching criterion for a single junction~\cite{10.1063/1.5144233}, we use
\begin{equation}
\varphi_{\mathrm{th}}(\tau)
=\pi-\arcsin[\frac{i_{\mathrm{b}}(\tau)}{2}]
\label{eq:escape-threshold}
\end{equation}
as a common phase threshold for voltage switching. We record the switching time $\tau_{\mathrm{sw}}$ when either $\varphi_1>\varphi_{\mathrm{th}}$ or $\varphi_2>\varphi_{\mathrm{th}}$ first occurs. The corresponding switching current is $i_{\mathrm{sw}}=v\tau_{\mathrm{sw}}$. The set of currents $i_{\mathrm{sw}}^{(j)}$ for $j=1,2,3,\ldots,N$ forms the SCD for the specified conditions. We use $N=10\,000$ realizations.

\begin{figure}[htbp]
\centering
\includegraphics[width=1\linewidth]{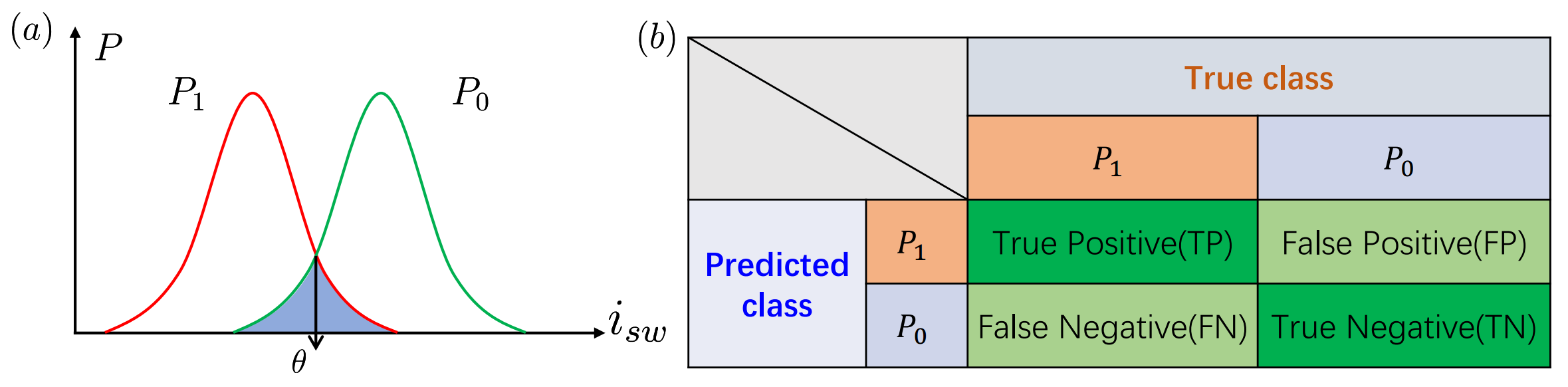}
\caption{ROC-AUC analysis of switching-current distributions. (a) Switching-current distributions with and without a microwave input. (b) Confusion matrix for binary classification.}
\label{fig:AUC}
\end{figure}

We denote the simulated SCDs without and with the microwave signal by
$P_0$ and $P_1$, respectively [Fig.~\ref{fig:AUC}(a)]. We quantify their separability by the area under the curve (AUC),  for the receiver operating characteristic (ROC) curve. Figure~\ref{fig:AUC}(b) shows the confusion matrix underlying the ROC-AUC ($\mathcal R_{AUC}$) calculation. An $\mathcal R_{AUC}$ of $0.5$ corresponds to chance-level discrimination, and an $\mathcal R_{AUC}$ of $1$ corresponds to complete separation. Following our earlier work~\cite{chai2026measuring}, we classify the signal as detectable when $\mathcal R_{AUC}\geq0.7$. The implementation and rationale for the method are described in Ref.~\cite{chai2026measuring}.
\section{Numerical Results and Physical Interpretation}
We use the model and statistical analysis above to examine how flux drive affects the detection sensitivity of a DC-SQUID to weak microwave signals.
\subsection{Effect of flux drive on weak-signal discrimination}
\begin{figure}[htbp]
\centering
\includegraphics[width=1\linewidth]{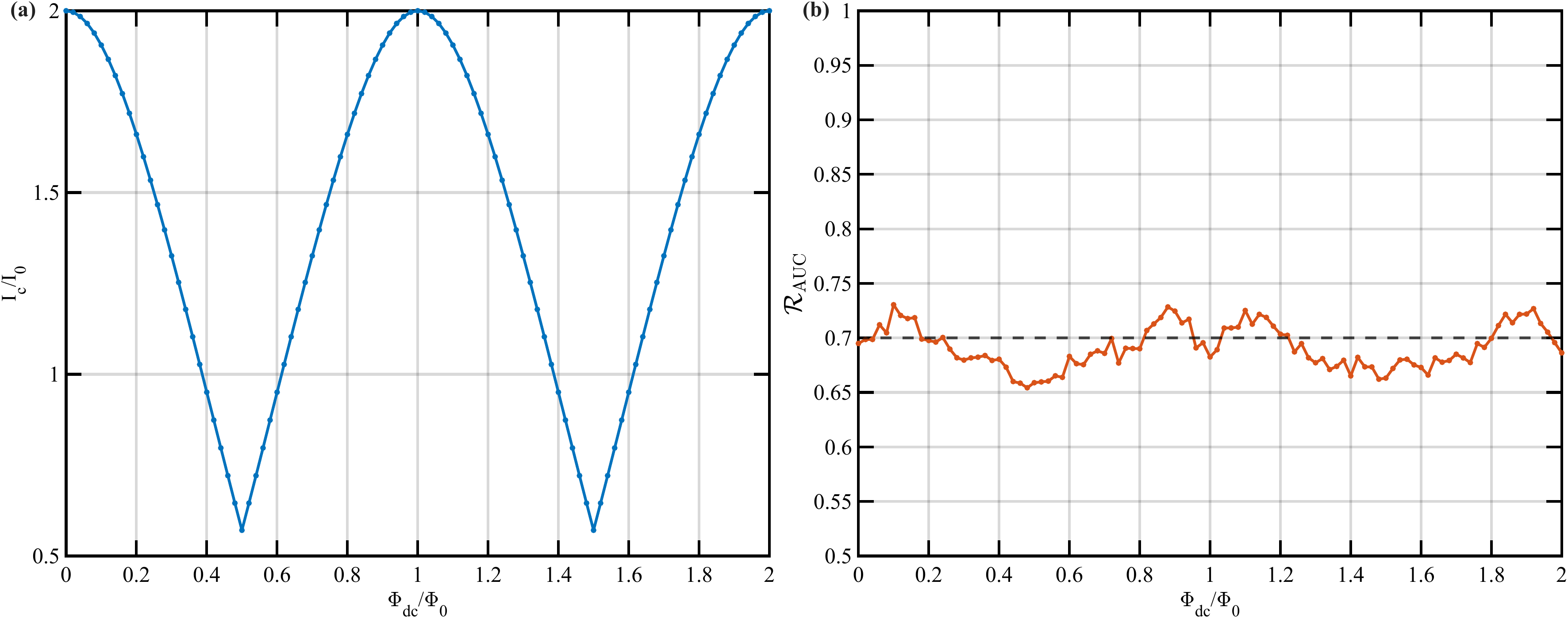}
\caption{(a) DC-SQUID critical current versus DC flux bias. (b) Simulated microwave detection sensitivity versus DC flux bias. The microwave frequency and power are $2.63$~GHz and $-120$~dBm, respectively.}
\label{fig:dc-flux-sensitivity}
\end{figure}

A DC flux bias modulates the critical current of a DC-SQUID, as shown in Fig.~\ref{fig:dc-flux-sensitivity}(a). We first assess whether this modulation alone improves sensitivity. For each DC flux $f_{\mathrm{dc}}=\Phi_{\mathrm{dc}}/\Phi_0$, we solve Eq.~\eqref{eq:RCSJ-normalized} with and without the microwave signal and compute $\mathcal R_{AUC}$ from the resulting SCDs.
Figure~\ref{fig:dc-flux-sensitivity}(a) shows that the critical current reaches minima near half-integer flux quanta, with a period of approximately one flux quantum. At a fixed input of $-120$~dBm, $\mathcal R_{AUC}$ varies only from about $0.65$ to $0.73$ [Fig.~\ref{fig:dc-flux-sensitivity}(b)], indicating little sensitivity gain from tuning $I_c$.

We next set $f_{\mathrm{dc}}=0$ and study the high-frequency alternating-current (AC) flux drive. Equation~\eqref{eq:mean-phase-equation} in Sec.~II\,B identifies the drive amplitude and frequency as key parameters. We fix the microwave frequency and power at $2.63$~GHz and $-125$~dBm, respectively. At each drive frequency and amplitude, we compute the SCDs with and without microwave input and evaluate their separability using $\mathcal R_{AUC}$. We set both initial phases to zero and sweep the drive frequency independently of the microwave frequency.
\begin{figure}[htbp]
\centering
\includegraphics[width=1\linewidth]{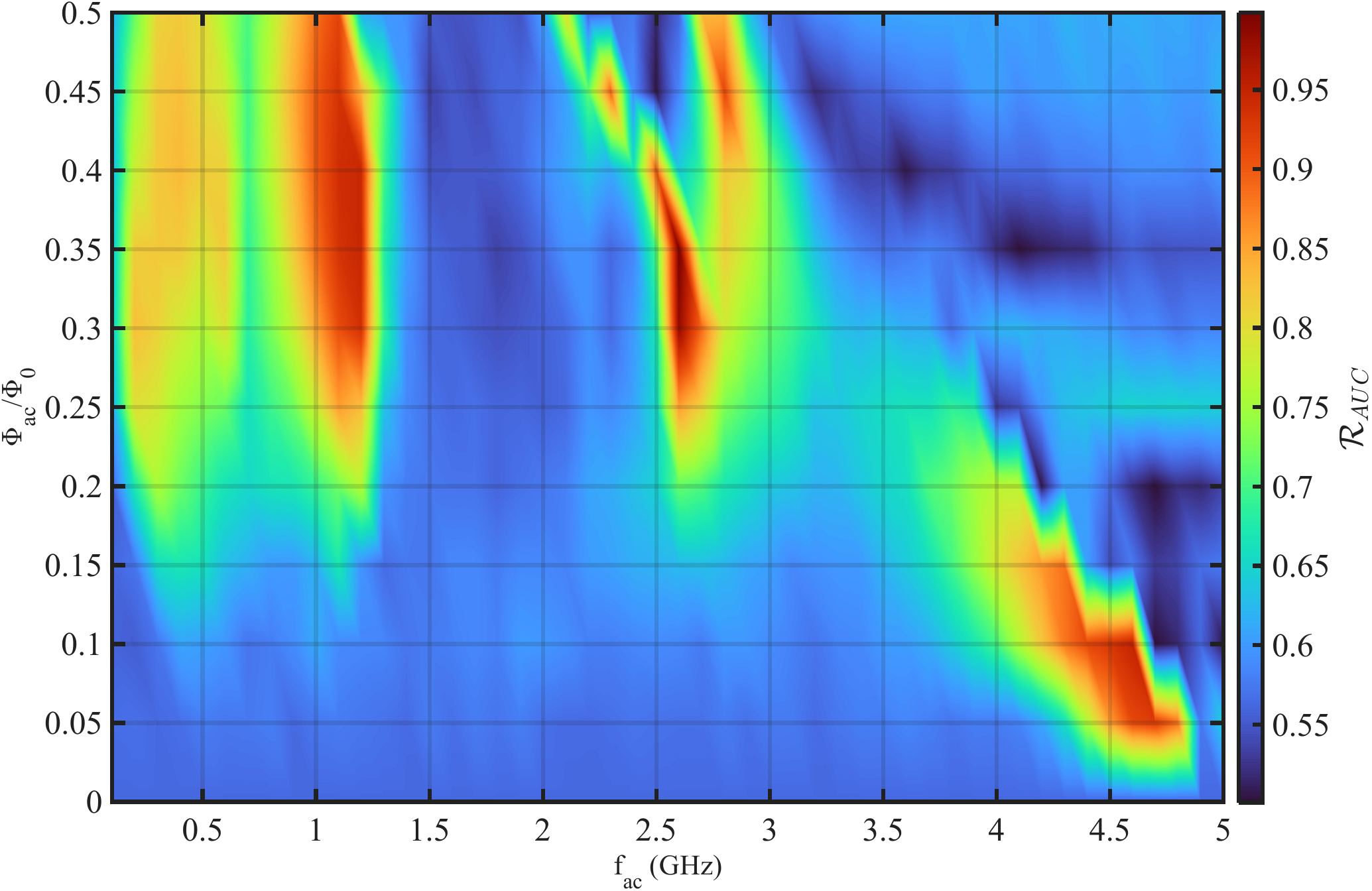}
\caption{DC-SQUID $\mathcal R_{AUC}$ versus the frequency and amplitude of the high-frequency flux drive. The microwave frequency is $2.63~\mathrm{GHz}$, the input power is $-125~\mathrm{dBm}$, and the static flux bias is zero.}
\label{fig:ac-flux-map}
\end{figure}

Figure~\ref{fig:ac-flux-map} shows localized regions of high $\mathcal R_{AUC}$ in the drive-frequency--amplitude plane. At $\Phi_{\mathrm{ac}}=0$, $\mathcal R_{AUC}$ is about $0.6$. With AC flux drive, several frequency and amplitude combinations yield $\mathcal R_{AUC}>0.9$. Because the device parameters, noise intensity, and microwave input are fixed, the variation reflects the effect of flux drive on the separation between the two SCDs. The high-$\mathcal R_{AUC}$ regions vary nonmonotonically with flux amplitude, reflecting the interplay of forced phase motion, microwave excitation, and nonlinear dynamics. For the following analysis, we choose $f_{\mathrm{ac}}=2.63~\mathrm{GHz}$ and $\Phi_{\mathrm{ac}}=0.35\Phi_0$ to maximize the microwave response.

\begin{figure}[htbp]
\centering
\includegraphics[width=1\linewidth]{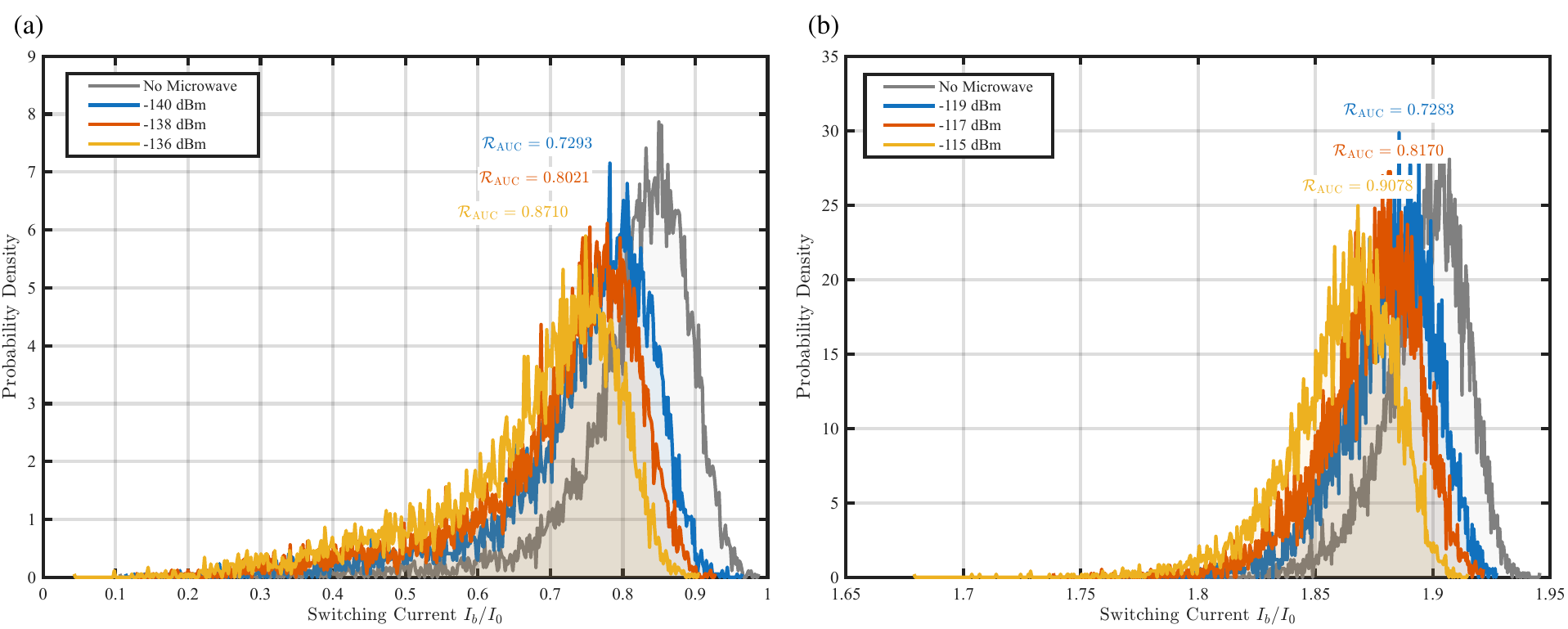}
\caption{Switching-current distributions at different continuous-wave microwave powers (a) with and (b) without high-frequency flux drive.}
\label{fig:WM}
\end{figure}
Figure~\ref{fig:WM} compares the discrimination performance with and without high-frequency flux drive. Without the drive, a $-119$~dBm signal yields $\mathcal R_{AUC}=0.7283$, above the detection threshold of $0.7$. At $-117$~dBm and $-115$~dBm, $\mathcal R_{AUC}$ increases to $0.8170$ and $0.9078$, respectively.
With high-frequency flux drive,
a $-140$~dBm signal yields $\mathcal R_{AUC}=0.7293$. At $-138~\mathrm{dBm}$ and $-136~\mathrm{dBm}$, the $\mathcal R_{AUC}$ reaches $0.8021$ and $0.8710$, respectively. The drive substantially improves microwave detection sensitivity.
\subsection{Reduced influence of thermal noise under high-frequency flux drive}
\label{subsec:temperature-response}
The simulations show that high-frequency flux drive sustains nonequilibrium phase dynamics and enhances weak-signal sensitivity. Our single-junction study~\cite{chai2026measuring} attributed the enhancement in the nonequilibrium regime to a reduced effect of thermal noise. We test whether the same mechanism applies to the DC-SQUID.
With no microwave input, we compare the SCDs at $T=30$, $35$, $40$, and $45~\mathrm{mK}$ with and without high-frequency flux drive.

\begin{figure}[htbp]
\centering
\includegraphics[width=1\linewidth]{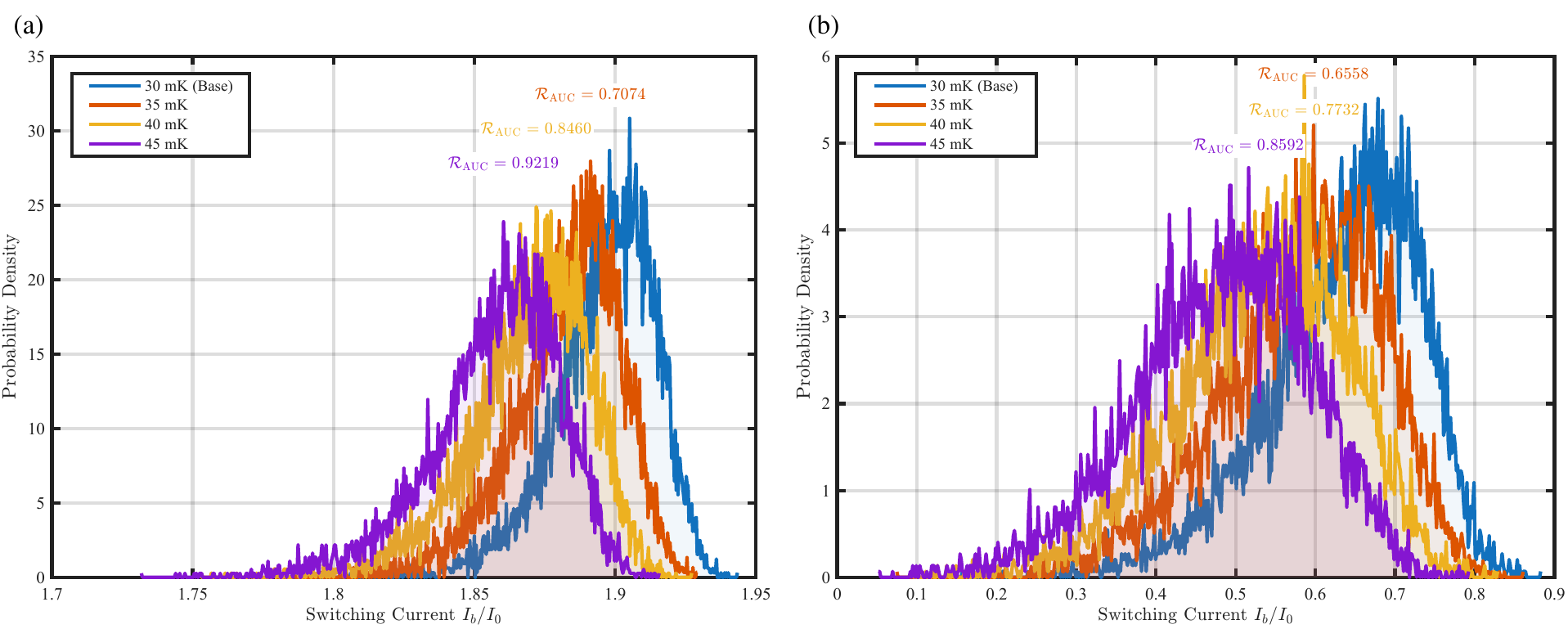}
\caption{DC-SQUID switching-current distributions at different temperatures (a) without and (b) with AC flux drive. Each $\mathcal R_{AUC}$ quantifies the separation between the SCD at the given temperature and that at $30~\mathrm{mK}$. The remaining simulation settings are as in Fig.~\ref{fig:ac-flux-map}.}
\label{fig:T}
\end{figure}
Without flux drive [Fig.~\ref{fig:T}(a)], increasing the temperature shifts the SCD toward lower currents and broadens its low-current tail. For $35$, $40$, and $45~\mathrm{mK}$ relative to $30~\mathrm{mK}$, the $\mathcal R_{AUC}$ are $0.7074$, $0.8460$, and $0.9219$, respectively, indicating a strong thermal effect. With high-frequency flux drive [Fig.~\ref{fig:T}(b)], the distribution shifts toward lower currents with increasing temperature, with a noticeably smaller displacement. The $\mathcal R_{AUC}$ at
$35$, $40$, and $45~\mathrm{mK}$ relative to $30~\mathrm{mK}$ are $0.6558$, $0.7732$, and $0.8592$, respectively. As in the single-junction case, the driven device is less sensitive to thermal noise.

Without flux drive, thermal activation dominates phase-particle escape. The escape rate increases exponentially with temperature, making the SCD strongly temperature dependent. Under high-frequency flux drive, both the phase particle and the potential barrier undergo sustained oscillations. The forced phase dynamics then largely determines escape, with thermal noise acting as a smaller perturbation. For the same temperature increase, the resulting SCDs are less distinguishable, as indicated by their lower $\mathcal R_{AUC}$.
\subsection{Dependence of detection sensitivity on relative phase}
\label{subsec:microwave-sensitivity}
The previous simulations set the initial phases of both the flux drive and the microwave signal to zero. We now examine how their relative phase affects the DC-SQUID sensitivity.

For a microwave signal at $-135~\mathrm{dBm}$ and $f_{\mathrm{rf}}=2.63~\mathrm{GHz}$, we vary the relative phase $\Delta\phi=\psi_{\mathrm{ac}}-\psi_{\mathrm{rf}}$ and calculate $\mathcal R_{AUC}$ from the SCDs with and without the microwave signal at each setting.
\begin{figure}[htbp]
\centering
\includegraphics[width=1\linewidth]{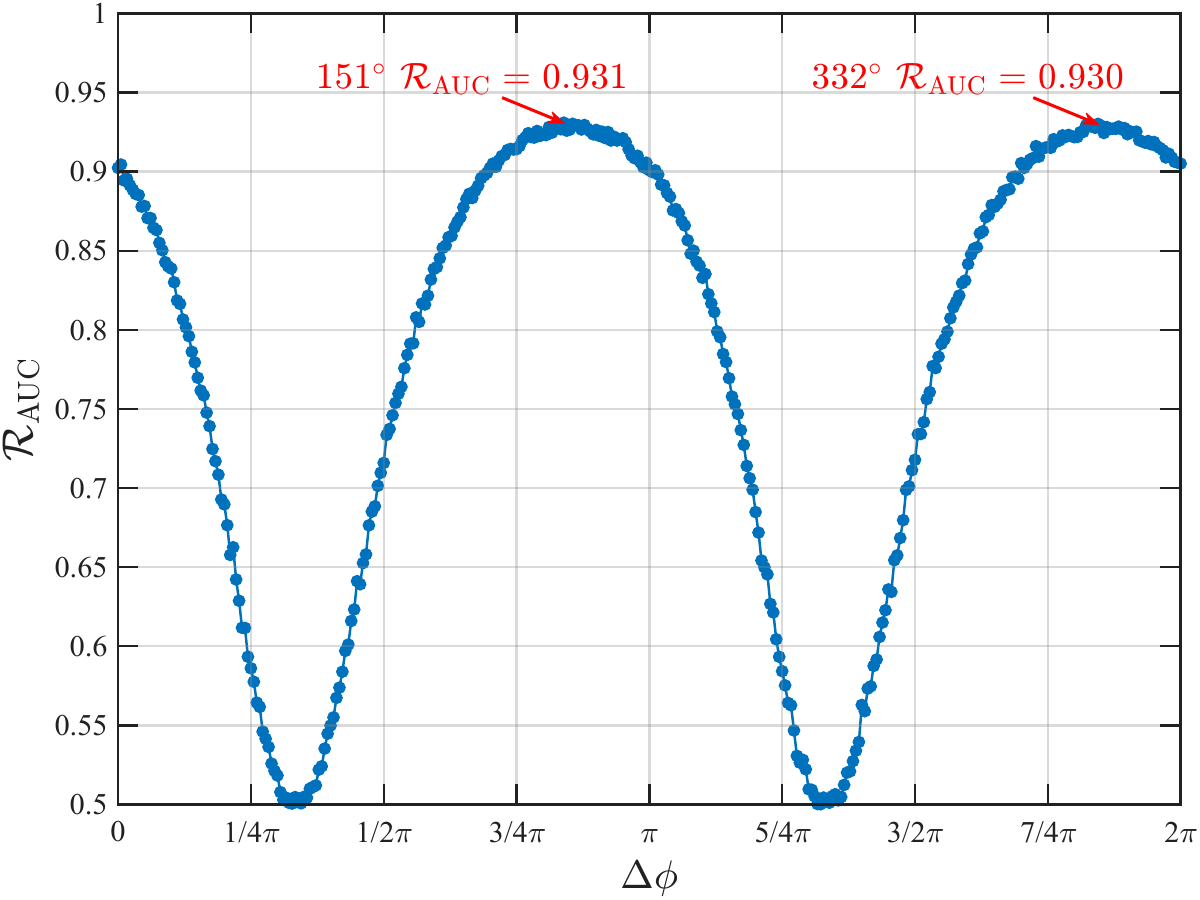}
\caption{Simulated $\mathcal R_{AUC}$ for a fixed microwave signal versus the relative phase $\Delta\phi$ between the flux drive and microwave signal. The microwave power is $P_{\mathrm{rf}}=-135~\mathrm{dBm}$, the frequencies are $f_{\mathrm{rf}}=f_{\mathrm{ac}}=2.63~\mathrm{GHz}$, and the flux amplitude is $\Phi_{\mathrm{ac}}=0.35\Phi_0$.}
\label{fig:phase-auc}
\end{figure}
Figure~\ref{fig:phase-auc} shows that $\mathcal R_{AUC}$ varies periodically with $\Delta\phi$, with peaks of $0.931$ and $0.930$ near $151^\circ$ and $332^\circ$, respectively, and minima close to $0.5$. The relative phase strongly influences sensitivity to the same signal. For a weak microwave signal of known frequency and phase, maximum sensitivity is expected when the flux drive is frequency matched and the phase difference is held near an $\mathcal R_{AUC}$ maximum.

\begin{figure}[htbp]
\centering
\includegraphics[width=1\linewidth]{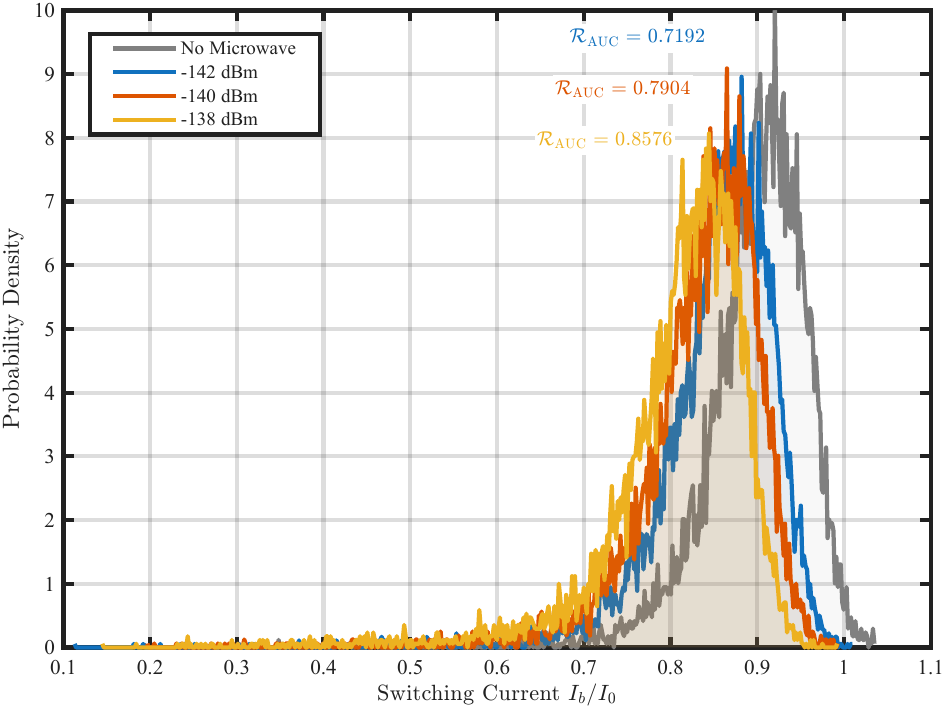}
\caption{Simulated DC-SQUID switching-current distributions and $\mathcal R_{AUC}$ at different microwave input powers. The relative phase is $\Delta\phi=151^\circ$. The remaining settings are as in Fig.~\ref{fig:phase-auc}.}
\label{fig:optimal-phase-power-response}
\end{figure}

With the settings of Fig.~\ref{fig:phase-auc}, we simulate SCDs over a range of microwave input powers and evaluate $\mathcal R_{AUC}$, as shown in Fig.~\ref{fig:optimal-phase-power-response}.
At $-142~\mathrm{dBm}$, the microwave signal remains detectable, with $\mathcal R_{AUC}=0.7192>0.7$. The minimum detectable power is therefore estimated as $P_{\min}\simeq-142~\mathrm{dBm}$, approximately $23~\mathrm{dB}$ below the undriven value in Fig.~\ref{fig:WM}.

To facilitate comparison with other detectors, we express this minimum power in bandwidth-normalized form. With a maximum bias current $I_{b,\max}=2.02I_0$ and sweep rate $v_I=0.053~\mathrm{A/s}$, the duration of one sweep is
\begin{equation}
T_{\mathrm r}=\frac{I_{b,\max}}{v_I}
\simeq5.34~\mu\mathrm{s}.
\end{equation}
Taking this duration as a uniform integration time, we estimate the one-sided equivalent noise bandwidth as
\begin{equation}
\Delta f_{\mathrm{eff}}\sim\frac{1}{2T_{\mathrm r}}
\simeq9.4\times10^4~\mathrm{Hz}.
\end{equation}
Using this estimated bandwidth $\Delta f_{\mathrm{eff}}$ and the minimum power satisfying the detection criterion, $P_{\min}=-142~\mathrm{dBm}\simeq6.31\times10^{-18}~\mathrm W$, we obtain
\begin{equation}
\mathcal P_{\mathrm{norm}}\equiv
\frac{P_{\min}}{\sqrt{\Delta f_{\mathrm{eff}}}}
\simeq2\times10^{-20}~\mathrm{W/\sqrt{Hz}}.
\label{eq:effective-NEP}
\end{equation}
Here $\mathcal P_{\mathrm{norm}}$ is the minimum detectable power normalized by the square root of the estimated bandwidth.
\subsection{Threshold response and discrimination of few-photon microwave pulses}
\label{subsec:pulsed-photon-response}
We next examine the DC-SQUID response to weak microwave pulses with energies corresponding to a few photons.
Following our earlier single-junction study~\cite{chai2026measuring}, we add a Gaussian-envelope microwave pulse to the total SQUID bias current,
\begin{equation}
\begin{aligned}
I_{\mathrm{rf}}^{(N_{\mathrm{ph}})}(t)
&=\sqrt{N_{\mathrm{ph}}}\,I_{1\mathrm{ph}}
\exp\!\left[-\frac{(t-t_d)^2}{2\sigma_t^2}\right]\\
&\quad\times\cos\!\left[2\pi f_{\mathrm{rf}}(t-t_d)\right],
\end{aligned}
\label{eq:photon-pulse-current}
\end{equation}
Here $N_{\rm ph}$ is the photon number, $t_d$ is the pulse center time, $\sigma_t$ is the Gaussian width of the current envelope, and $I_{1\mathrm{ph}}$ is the peak current corresponding to the energy of one photon.
In the simulations, $I_{\mathrm{rf}}^{(N_{\mathrm{ph}})}(t)/I_0$ replaces the continuous-wave term in Eq.~\eqref{eq:normalized-current}. We assume complete current coupling to the SQUID and neglect impedance mismatch and transmission loss. The pulse energy is
\begin{equation}
E_{\mathrm{p}}
=Z_0\int_{-\infty}^{\infty}
\left[I_{\mathrm{rf}}^{(N_{\mathrm{ph}})}(t)\right]^2\,dt
=N_{\mathrm{ph}}h_{\mathrm{P}}f_{\mathrm{rf}},
\label{eq:photon-pulse-energy}
\end{equation}
where $h_{\mathrm{P}}$ is Planck's constant and $Z_0=50~\Omega$ is the port impedance. In the limit $2\pi f_{\mathrm{rf}}\sigma_t\gg1$, integrating the squared current in Eq.~\eqref{eq:photon-pulse-current} gives
\begin{equation}
I_{1\mathrm{ph}}\simeq
\sqrt{\frac{2h_{\mathrm{P}}f_{\mathrm{rf}}}
{Z_0\sqrt{\pi}\,\sigma_t}}.
\label{eq:single-photon-current-calibration}
\end{equation}
Thus $N_{\mathrm{ph}}$ is the equivalent photon number defined by the input pulse energy.

\begin{figure}[htbp]
\centering
\includegraphics[width=1\linewidth]{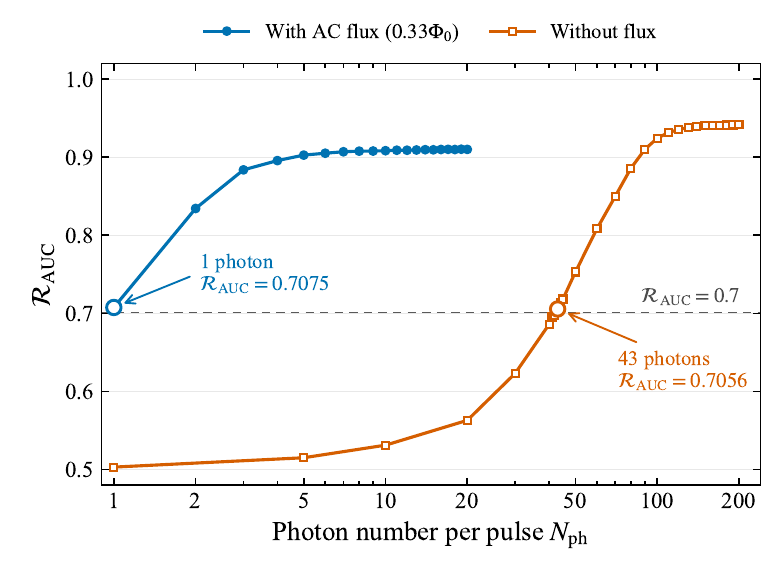}
\caption{DC-SQUID $\mathcal R_{AUC}$ versus equivalent photon number for microwave pulses. Blue circles denote high-frequency flux drive, and orange squares denote the undriven case. The horizontal dashed line marks the detection threshold $\mathcal R_{AUC}=0.7$. The pulse carrier frequency is $2.63~\mathrm{GHz}$ and the Gaussian width is $5~\mathrm{ns}$.}
\label{fig:photon-auc-flux-comparison}
\end{figure}

We use $f_{\mathrm{rf}}=2.63~\mathrm{GHz}$ and $\sigma_t=5~\mathrm{ns}$. The corresponding single-photon energy is $h_{\mathrm{P}}f_{\mathrm{rf}}=1.743\times10^{-24}~\mathrm{J}$, giving $I_{1\mathrm{ph}}\simeq2.805~\mathrm{nA}$ from Eq.~\eqref{eq:single-photon-current-calibration}. Guided by Sec.~III\,C, we set the high-frequency flux drive to
\begin{equation}
\Phi_{\mathrm{ext}}(t)
=0.33\Phi_0\cos\!\left[2\pi f_{\mathrm{rf}}(t-t_d)+150^\circ\right].
\label{eq:photon-continuous-flux}
\end{equation}
The DC flux is zero, the high-frequency drive is matched to the pulse carrier frequency, and the flux phase at the pulse center is $150^\circ$. For the undriven control, we set $\Phi_{\mathrm{ext}}=0$. The finite envelope and phase settings pertain to the pulsed protocol, for which the phase dependence differs from that of the continuous-wave case. The flux drive shifts the switching-current range, so the pulse timing differs from our previous single-junction protocol, which centered the pulse halfway through a bias sweep~\cite{chai2026measuring}. For each flux condition, we use the $N_{\mathrm{ph}}=0$ SCD as a reference and center the pulse near the onset of escape on its low-current side. With sweep duration $T_{\mathrm{r}}=I_{b,\max}/\dot I_b$, we use
\begin{equation}
\frac{t_d}{T_{\mathrm{r}}}=
\begin{cases}
0.495, & \Phi_{\mathrm{ac}}=0.33\Phi_0,\\
0.9249, & \Phi_{\mathrm{ac}}=0.
\end{cases}
\label{eq:photon-pulse-timing}
\end{equation}
For $I_{b,\max}=2.02I_0$, $\dot I_b=0.053~\mathrm{A/s}$, and $T_{\mathrm{r}}\simeq5.336~\mu\mathrm{s}$, these settings give $t_d\simeq2.641~\mu\mathrm{s}$ and $4.935~\mu\mathrm{s}$, respectively.
The corresponding bias currents at the pulse centers are $139.986~\mathrm{nA}$ and $261.555~\mathrm{nA}$. The simulated cumulative switching fractions at those currents are $9.94\%$ and $6.80\%$, respectively. Thus, both pulse centers lie on the low-current side of the background distribution, while most trials are still in the zero-voltage state.
At a bath temperature of $T=30~\mathrm{mK}$, we calculate $\mathcal R_{AUC}$ between each pulse SCD and its background SCD for both flux conditions, as shown in Fig.~\ref{fig:photon-auc-flux-comparison}. Without high-frequency flux drive, $\mathcal R_{AUC}$ increases with pulse energy. For $N_{\mathrm{ph}}=40$, $42$, and $43$, the $\mathcal R_{AUC}$ are $0.6859$, $0.6972$, and $0.7056$, respectively, giving a minimum detectable equivalent photon number of $43$. With high-frequency flux drive, a pulse with the energy of one photon yields $\mathcal R_{AUC}=0.7075$, slightly above the nominal detection threshold of $0.7$. For $N_{\mathrm{ph}}=2$ and $4$, the corresponding AUCs are $0.8343$ and $0.8957$, respectively.
In addition, Fig.~\ref{fig:photon-auc-flux-comparison} also shows a range over which different equivalent photon numbers produce distinguishable responses. The high-frequency flux drive improves sensitivity but narrows this dynamic range. The $\mathcal R_{AUC}$ varies approximately linearly over $N_{\mathrm{ph}}\in[43,110]$ without drive and over $N_{\mathrm{ph}}\in[1,4]$ with drive.

\section{Conclusions and Discussion}
Building on our numerical study of weak microwave detection using nonequilibrium dynamics in a single JJ, we investigated the feasibility of driving a DC-SQUID out of equilibrium with high-frequency magnetic flux to improve weak-signal detection.
As in the single-junction case, the nonequilibrium DC-SQUID is less sensitive to thermal noise, allowing it to approach the single-microwave-photon energy scale. Under the adopted binary-classification criterion, the simulations show statistical discrimination at the single-photon energy scale. The detector also shows a photon-number-dependent response over a narrow dynamic range.
The proposed protocol distinguishes two switching-current distributions obtained from repeated trials. Its demonstrated sensitivity therefore applies to ensemble statistics, and single-shot detection remains beyond the scope of the protocol.
The flux-driven DC-SQUID is more experimentally accessible than the nonequilibrium single-junction scheme because it can operate at moderate $Q$ and bias-current sweep rates.
For phase-sensitive operation, the high-frequency flux drive must be synchronized with the microwave signal.

The simulations include thermal noise and describe idealized operating conditions. Macroscopic quantum tunneling, external flux fluctuations, mismatches in the junction critical currents, resistances, and capacitances, and asymmetric drive coupling may also affect detection sensitivity and need to be considered experimentally. The flux-drive amplitude and frequency, bias-current sweep rate, and microwave pulse injection timing and coupling should be chosen to maximize sensitivity.

\section{Acknowledgments}
This work is partially supported by the Central Guidance Funds for Local Science and Technology Development (No. 2025ZYD0154), and the National Key Research and Development Program of China (NKRDC) under Grant No. 2021YFA0718803.
\bibliography{Ref}
\end{document}